\documentclass[prl,twocolumn,showpacs]{revtex4}

\usepackage{graphicx}% Include figure files
\usepackage{dcolumn}% Align table columns on decimal point
\usepackage{bm}% bold math
\usepackage{amsmath}
  
\begin{document}
\title{Trends in Elasticity and Electronic Structure of Transition-Metal Nitrides and Carbides from First Principles}
\author{Zhigang Wu}
\affiliation{Geophysical Laboratory, Carnegie Institution of Washington, Washington, DC 20015}
\author{Xiao-Jia Chen}
\affiliation{Geophysical Laboratory, Carnegie Institution of Washington, Washington, DC 20015}
\author{Viktor V. Struzhkin}
\affiliation{Geophysical Laboratory, Carnegie Institution of Washington, Washington, DC 20015}
\author{Ronald E. Cohen}
\affiliation{Geophysical Laboratory, Carnegie Institution of Washington, Washington, DC 20015}

\date{\today}

\begin{abstract}
The elastic properties of the $B_1$-structured transition-metal nitrides and their carbide counterparts are studied using 
the {\it ab initio\ } density functional perturbation theory. The linear response results of elastic constants are in 
excellent agreement with those obtained from numerical derivative methods, and are also consistent with measured data. We 
find the following trends: (1) Bulk moduli $B$ and tetragonal shear moduli $G^{\prime}=(C_{11}-C_{12})/2$, increase and 
lattice constants $a_{0}$ decrease rightward or downward on the Periodic Table for the metal component or if C is replaced 
by N; (2) The inequality $B > G^{\prime} > G > 0$ holds for  $G=C_{44}$; (3) $G$ depends strongly on the number of valence 
electrons per unit cell ($Z_{V}$). From the fitted curve of $G$ as a function of $Z_{V}$, we can predict that MoN is unstable 
in $B_{1}$ structure, and transition-metal carbonitrides ($e.g.$ ZrC$_{x}$N$_{1-x}$) and di-transition-metal carbides 
($e.g.$ Hf$_{x}$Ta$_{1-x}$C) have maximum $G$ at $Z_{V} \approx 8.3$. 
\end{abstract}
\pacs{62.20.Dc, 71.15.Mb, 74.70.Ad}
\maketitle

Elasticity describes the response of a crystal under external strain and provides key information of the bonding 
strength between nearest-neighbor atoms. The information obtained from accurate calculation of elasticity is essential 
for understanding the macroscopic mechanical properties of solids and for the design of hard materials. Nowadays it is 
possible to calculate elasticity using $ab$ $initio$ quantum-mechanical techniques, and $ab$ $initio$ calculations have 
proven to be very powerful in not only providing accurate elastic constants or moduli in good agreement with measurements 
\cite{czha} but also predicting elasticity at extreme conditions of high temperatures and high pressures \cite{lars,gerd}, 
which are not easily accessible to experiment but have wide applications in the fields ranging from solid-state physics 
to seismology. Most previous $ab$ $initio$ calculations of elasticity used finite strain methods within the framework of 
the density-functional theory (DFT). The development of density-functional perturbation theory (DFPT) makes it possible 
now to obtain elastic constants directly and more accurately \cite{dfpt,hamann04}. 

Transition-metal nitrides and carbides in the rocksalt ($B_1$) structure are widely used for cutting tools, magnetic 
storage devices, generators and maglev trains due to their high hardness, high melting points and oxidation 
resistance \cite{toth71}.
These excellent properties are associated with their unusual electronic bonding. The relatively high superconducting 
transition temperature in some of these compounds, reaching nearly 18 K in NbC$_{1-x}$N$_{x}$ \cite{matt}, indicates 
a strong electron-phonon interaction. Many theoretical studies of their electronic structure 
\cite{papa,chen88,grossman99,jhi99,freeman01,jhi01} have revealed an unusual mixture of covalent, metallic, and ionic 
contributions to bonding which must ultimately lie at the root of their unusual properties. Specifically, it was found 
\cite{jhi99} that the hardness enhancement of these materials can be understood on a fundamental level in terms of their 
electronic band structures. But the general trends of elasticity and electronic structure among the transition-metal 
nitrides and carbides remain unclear and a challenge for engineering hard materials.

In this Letter we perform systematic calculations of elasticity of the selected transition-metal 
nitrides and carbides within the framework of the density functional perturbation theory. The accuracy of this method is also 
examined by numerical finite strain methods. We obtain interesting trends of the variation of the bulk and shear moduli 
among various compounds. We demonstrate the electronic origin of such trends in these materials. 

Elastic constants are defined as 
\begin{equation}
c_{ijkl} \equiv \frac{\partial \sigma_{ij}} {\partial \epsilon_{kl}}
	      = \frac{1}{\Omega} \frac{\partial^2 E_{\rm el}}
                {\partial \epsilon_{ij} \partial \epsilon_{kl}} 
              - \delta_{ij} \sigma_{kl} = C_{ijkl} - \delta_{ij} \sigma_{kl}~~,
\label{eq1}
\end{equation}
where $\sigma$ is stress, $\epsilon$ is strain, $\Omega$ and $E_{\rm el}$ are the unit-cell volume and energy, $\delta$ 
is the Kronecker delta function, and the Latin indices run from 1 to 3. At zero hydrostatic pressure $c_{ijkl} = C_{ijkl}$. 
One can apply a small strain and calculate the change of energy or stress to obtain elastic constants. Direct calculations 
of stress are possible from the quantum mechanical theory of stress \cite{stress}. 

Alternatively elastic constants can be viewed as the linear response (LR) of an undisturbed system by homogeneous strains 
(macroscopic distortions of the crystal), and can be directly computed by the DFPT \cite{dfpt}. However, strain perturbation 
is much more difficult to cope with within the DFPT than atomic displacement perturbation (phonon) in that a homogeneous 
strain changes the boundary conditions of the Hamiltonian of an infinite system, and the DFPT requires the same basis set for 
the undisturbed and disturbed systems. 

Recently Hamann {\it et al.} \cite{hamann04} applied the reduced-coordinate metric tensor method to the LR of strain type 
perturbations. Their approach \cite{hamann04} is based on an energy expression in terms of reduced coordinates, which are 
defined in both real and reciprocal space using primitive direct ({\bf R}) and reciprocal ({\bf G}) lattice vectors. If the 
reduced coordinates are denoted with a tilde, a real space vector {\bf X} and a reciprocal space vector {\bf K} are given by
\begin{equation}
X_i = \sum_j R_{ij} {\tilde X}_j, \: K_i = \sum_j G_{ij} {\tilde G}_j~~,
\label{eq4}
\end{equation}
respectively. Essentially every energy term of the DFT can be expressed as dot products of vectors 
in real or reciprocal space, and
\begin{equation}
{\bf X}^{\prime} \cdot {\bf X} = \sum_{ij} {\tilde X}^{\prime}_i \varXi_{ij} {\tilde X}_j~, \:
{\bf K}^{\prime} \cdot {\bf K} = \sum_{ij} {\tilde K}^{\prime}_i \varUpsilon_{ij} {\tilde K}_j~~, 
\label{eq5}
\end{equation}
where the metric tensors ${\bm \varXi}$ and ${\bm \varUpsilon}$ are
\begin{equation}
\varXi_{ij} = \sum_{k} R_{ki} R_{kj}~, \: \varUpsilon_{ij} = \sum_{k} G_{ki} G_{kj}~. 
\label{eq6}
\end{equation}
The DFT energy derivatives with respect to strain act only on the metric tensors, and the first order derivatives are
\begin{equation}
\frac{\partial \varXi_{ij}}{\partial \epsilon_{kl}} = R_{ki}R_{lj} + R_{li}R_{kj}, \:
\frac{\partial \varUpsilon_{ij}}{\partial \epsilon_{kl}} = -G_{ki}G_{lj} - G_{li}G_{kj}.
\label{eq7}
\end{equation}
and second order derivatives can be directly obtained. After this transformation, strain has an equal footing with other coordinates, 
and the wave functions have invariant boundary conditions, so that the strain derivatives such as elastic constants can be evaluated 
in a way similar to other derivatives. Note that the DFPT expressions give  $C_{ijkl}$, not $c_{ijkl}$. Elastic constants were computed 
for an insulator AlP \cite{hamann04}, and the agreement between DFPT and numerical finite strain methods is perfect. Although the DFPT 
is directly applicable to metals, it needs special care for the Fermi surface due to partially occupied states. Here we examine the 
accuracy of DFPT for transition-metal nitrides and carbides. 

\begin{figure}[tbp]
\begin{center}
\includegraphics[width=\columnwidth]{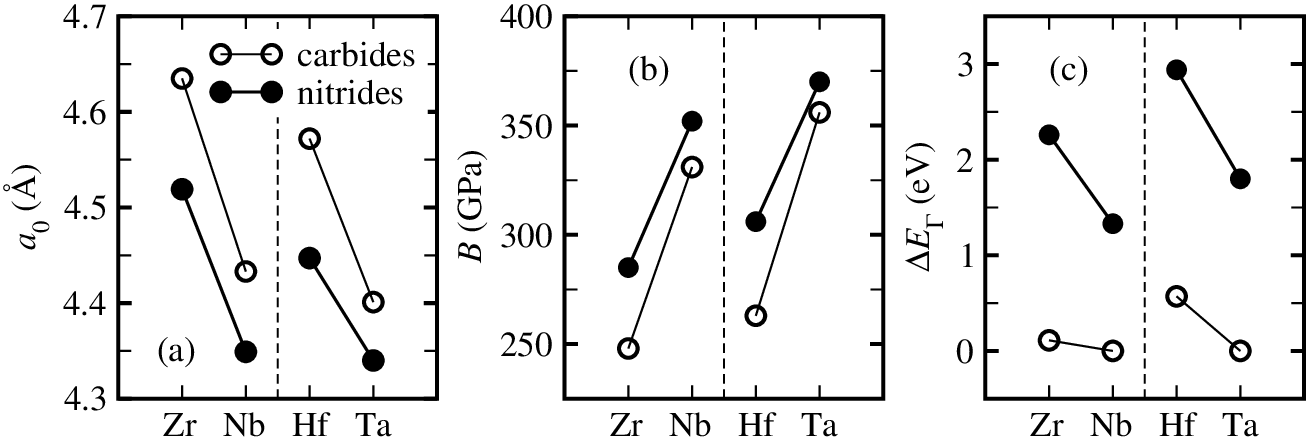}
\end{center}
\caption{\label{fig1} Theoretical (a) lattice constants $a_{0}$, (b) bulk moduli $B$, and (c) zone-center band gap $\Delta E_{\Gamma}$ 
for selected transition-metal nitrides and carbides with $B_{1}$ structure. Filled symbols denote nitrides, and open symbols refer to 
carbides.}
\end{figure}

Our first principles calculations are based on the DFT within the local density approximation (LDA). We used ABINIT (version 4.3.3) 
\cite{abinit}, which uses plane-wave basis sets and norm-conserving pseudopotentials. The pseudopotentials of C, N, Zr, Nb, Hf and Ta 
were generated using the OPIUM program \cite{opium}. The cutoff energy of plane-wave basis is 50 Ha. A dense $16 \times 16 \times 16$ 
${\bf k}$-point mesh was used over the Brillouin zone (BZ), and the cold smearing method \cite{marzari96} was performed for BZ integration.

Equilibrium lattice constants $a_0$ and bulk moduli $B$ were evaluated from fitting energy and volume data to the Vinet equation of state 
\cite{vinet}. Compared with the available experimental data \cite{toth71,chen04,weber73,brown66}, $a_0$ are about 1.0-1.5\% smaller, while 
in average $B$ are 15\% larger. These are typical LDA errors. The presence of N or C in transition-metal compounds leads to the hybridized 
$p$-$d$ bonding, and its strength determines lattice constants and bulk moduli. As illustrated in Fig. \ref{fig1}(a) and \ref{fig1}(b), 
$a_0$ decrease and $B$ increase as the transition-metal proceeds rightward (e.g. Zr $\rightarrow$ Nb) or downward (e.g. Zr $\rightarrow$ Hf) on 
the Periodic Table. Smaller $a_{0}$ and larger $B$ indicate stronger bonding. The former trend is because of more filling of the metal 
$d$-bands, which enhances the $d$-$p$ bonding \cite{grossman99}. The later trend is a result of the presence of $f$-electrons in core states, 
which repel $d$-orbitals out of core regions and in turn make them bond with C or N $p$-electrons more tightly. Available experimental data 
support these theoretical conclusions. Fig. \ref{fig1} also show that nitrides have smaller $a_0$ and larger $B$ than their corresponding 
carbides, because N has one more valence electron than does C so that the $d$-$p$ bonding strength in nitrides is higher than that in their 
carbide counterparts. Experiment and theory agree well with the trend of $a_0$, but the existing measured moduli of NbN and ZrN are lower 
than those of NbC and ZrC, respectively. Because different experimental methods have different level of accuracy, and experimental 
circumstances ($e.g.$ temperature, purity, etc) could affect measured results greatly, experimentally re-scrutinizing $B$ is needed to 
clarify the contradiction with theory.

\begin{table}[tbp]
\caption{Calculated bulk moduli and elastic constants (GPa) of NbN in $B_1$ structure. Three different approaches were used. See text for 
the meaning of theory 1, 2, and 3. 
\label{tab1}}
\begin{ruledtabular}
\begin{tabular}{l|cccc}
 method   & $B$    & $C_{11}$    & $C_{12}$    & $C_{44}$  \\      
\hline			     	     	   
 theory 1 & 353.6  & 742.6       &  159.1      &  76.4     \\
 theory 2 & 353.6  & 739.4       &  160.8      &  75.5     \\
 theory 3 & 353.7  & 738.9       &  161.1      &  74.8     \\
\end{tabular}
\end{ruledtabular}
\end{table}

\begin{figure}[tbp]
\begin{center}
\includegraphics[width=\columnwidth]{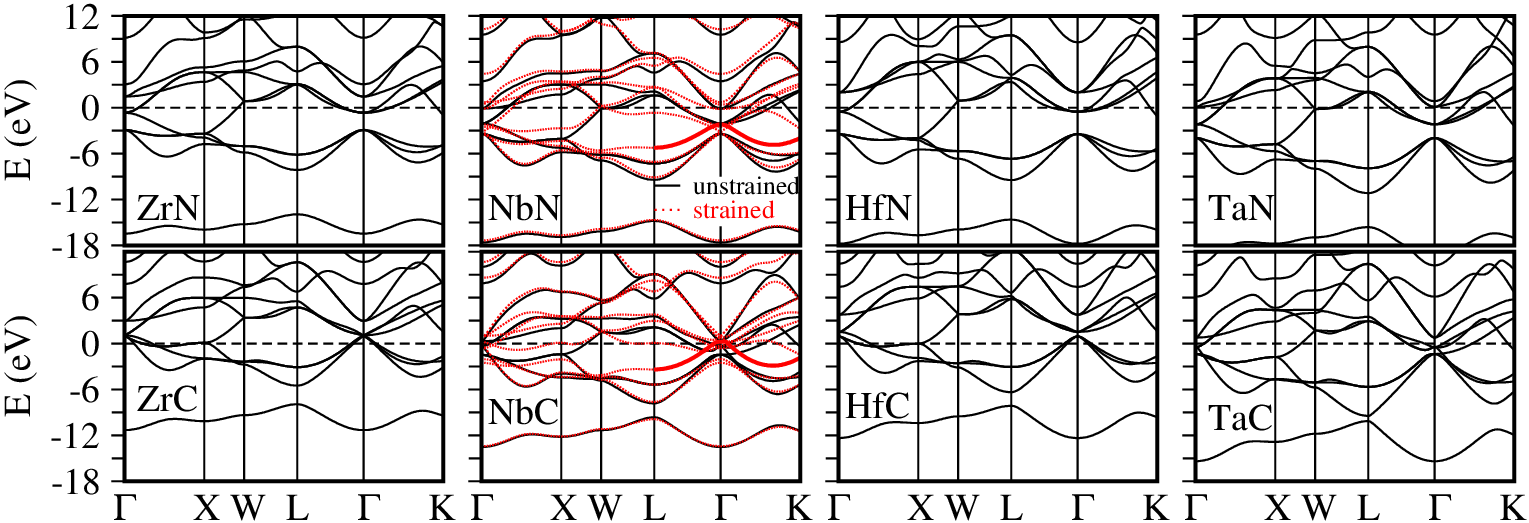}
\end{center}
\caption{\label{fig2} Electronic band structures of selected transition-metal nitrides and carbides with $B_{1}$ structure. The red curves 
are band structures of NbN and NbC under a finite shear strain ($\epsilon_{12} =  \epsilon_{21} = 0.1$).}
\end{figure}

Electronic band structures were calculated at zero pressure, and it can be seen in Fig. \ref{fig2} that all these nitrides and carbides 
in $B_{1}$ structure are metals, since there is a direct band overlap at the X point. At the zone-center ($\Gamma$-point), there is a 
finite band gap $\Delta E_{\Gamma}$, except for NbC and TaC. $\Delta E_{\Gamma}$ (Fig. \ref{fig1}(c)), the $d$-band splitting breadth, 
increases as the transition-metal goes leftward on the Periodic Table due to an upward shift of the metal $d$-bands as a result of less 
filling \cite{freeman01}, or goes downward because the presence of $f$-electrons promotes metal $d$-band energies. Carbides have smaller 
$\Delta E_{\Gamma}$ than their nitride counterparts since the carbon $p$-bands have higher energy than nitrogen.

Elastic constants were computed in the Voigt notation, in which the Greek subscripts of $C_{\mu \nu}$ run from 1 to 6. Crystals in cubic 
structure have only three independent elastic constants, namely $C_{11}$, $C_{12}$, and $C_{44}$, and the bulk modulus is
\begin{equation}
B = \frac{1}{3} (C_{11}+2C_{12})~~.
\label{eq10}
\end{equation}
One can apply the following strains
\begin{equation*}
\left.   {\bm \epsilon}_{\rm tetr} = \frac{1}{3} \begin{pmatrix} 
         -\delta & 0       & 0        \\ 
         0       & -\delta & 0        \\
         0       & 0       & 2\delta  \end{pmatrix}~~, \:
\right.  {\bm \epsilon}_{\rm orth} = \begin{pmatrix} 
         0       & \delta  & 0        \\ 
         \delta  & 0       & 0        \\
         0       & 0       & \delta^2 \end{pmatrix}~~, 
\end{equation*}
to distort the lattice vectors, ${\bf R}^{\prime} = (1+{\bm \epsilon}) {\bf R}$. The resulting changes of energy density 
($U=E_{\rm el}/\Omega$) are associated with elastic constants, 
\begin{align}
U_{\rm tetr} &= \frac{1}{3}(C_{11}-C_{12})\delta^2 + O(\delta^3)~~, \label{eq11} \\
U_{\rm orth} &= 2C_{44}\delta^2 + O(\delta^4)~~, \label{eq12}
\end{align}
respectively. $C_{11}$ and $C_{12}$ can be obtained from Eqs. (\ref{eq10}) and (\ref{eq11}), and $C_{44}$ from Eq. (\ref{eq12}). We denote 
this finite strain approach based on total energy as ``theory 1''. Another way is to compute the stress tensor elements for selected strains. 
For $C_{11}$ and $C_{12}$ we applied a strain tensor with $\epsilon_{11} = \delta$ (zero for other elements), and $C_{11} = 
\sigma_{11} / \delta$, $C_{12} = \sigma_{22} / \delta$. For $C_{44}$ the applied strain was $\epsilon_{12} = \epsilon_{21} = \delta$, and 
$C_{44} = \sigma_{12} / 2\delta$. This is called ``theory 2''. We used $\delta = \pm 0.002$, $\pm 0.004$. The third method (theory 3) 
is the DFPT of LR of strain perturbation \cite{dfpt,hamann04}. For comparison, elastic constants of NbN obtained from these three methods 
are shown in Table \ref{tab1}, where $B$ were calculated from $C_{11}$ and $C_{12}$ by Eq. (\ref{eq10}) for theory 2 and 3. High level of 
agreement between the DFPT and the energy (theory 1) or stress (theory2) was achieved, and similar agreement was also made for other 
compounds. It confirms the reliability of the LR theory of strain perturbation for elastic constants of metals.

As summarized in Table \ref{tab2}, present DFPT elastic constants are in accordance with experimental results. We point out that the 
uncertainty in neutron scattering measurements of elastic constants could be as high as 10-15 \%, and the experimental data in Table 
\ref{tab2} were measured at room temperature. The elastic stability criteria for a cubic crystal \cite{wang93} at ambient conditions are
\begin{equation}
C_{11}+2C_{12} > 0,~ C_{44} > 0~~,~ {\rm and}~ C_{11}-C_{12} > 0~~,
\label{eq13}
\end{equation}
$i.e.$ all the bulk $B$, shear $G=C_{44}$, and tetragonal shear $G^{\prime}=(C_{11}-C_{12})/2$ moduli are positive.
Our results satisfy all three criteria in Eq. (\ref{eq13}), and it follows that these materials in rocksalt structure are stable, 
consistent with experiment. In general $ B > G > G^{\prime}$, but for these materials both theory and experiment hold that $ B > G^{\prime} 
> G > 0$, so the shear modulus $G$ is the main constraint on stability.

%\begin{widetext}
%\end{widetext}

\begin{figure}[tbp]
\begin{center}
\includegraphics[width=\columnwidth]{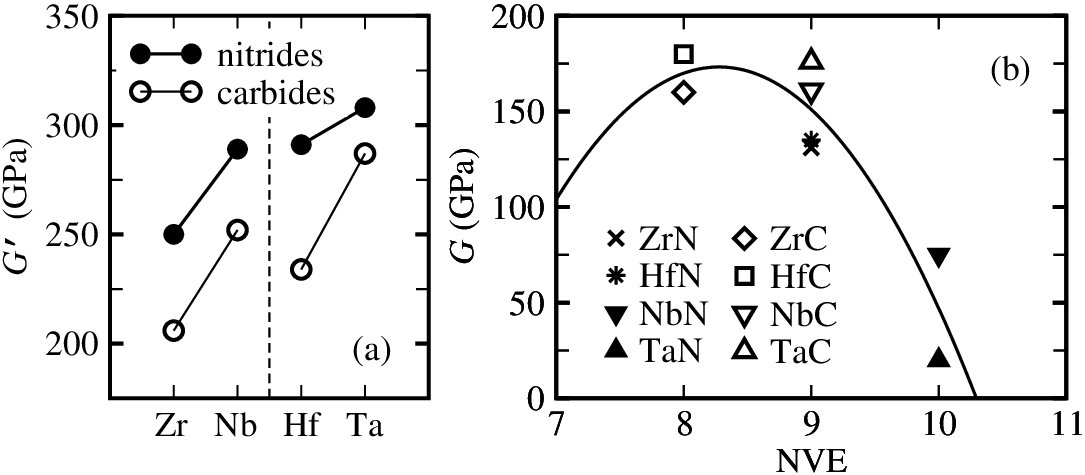}
\end{center}
\caption{\label{fig3} Theoretical (a) tetragonal shear moduli $G^{\prime}=(C_{11}-C_{12})/2$, and (b) shear moduli $G=C_{44}$ with respect 
to number of valence electrons per unit cell ($Z_{V}$). The solid curve in (b) is obtained by quadratic fitting to average shear moduli at 
each $Z_{V}$.}
\end{figure}

The tetragonal shear modulus $G^{\prime}$ measures the response of a crystal under volume conserving tetragonal shear strain, which involves 
stretching of metal-N or -C bonds and bending of metal-metal bonds. Because in these crystals $C_{11} \gg C_{12}$, and $C_{11}$ is determined 
by the nearest-neighbor interaction, similar to bulk modulus $B$, $G^{\prime}$ have the same trends as $B$, as seen in Fig. \ref{fig3}(a) 
that (1) $G^{\prime}$ increases as the metal goes rightward or downward on the Periodic Table; (2) Nitrides have bigger $G^{\prime}$ than their 
carbide counterparts. These trends generally are supported by experiment, and they are a result of the enhancement of $d$-$p$ bonding due to 
adding valence electrons or the presence of $f$-electrons, as discussed in previous paragraphs for $B$. However, farther right on the 
Periodic Table ($e.g.$ Nb $\rightarrow$ Mo) valence electrons saturate the bonding states and begin to fill the antibonding states,
so $B$ and $G^{\prime}$ will go down after reaching a peak.

The shear modulus $G$ ($C_{44}$) is the most important parameter governing indentation hardness. The hardness of a material is defined as 
its resistance to another material penetrating its surface, and it is determined by the mobility of dislocations. In covalent hard materials, 
the bond-breaking energy under plastic deformation and the bond-restoring energy under elastic shear strain are very similar, so that 
one of the determining factors of hardness is the response of covalent bonds to shear strain \cite{jhi99}. We classified these materials 
according to their number of valence electrons per unit cell ($Z_{V}$) in Fig. \ref{fig3}(b), which shows that $G$ is intimately related to 
$Z_{V}$. At $Z_{V} = 8$ and 9, these crystals have large $G$, but a modest decrease occurs when $Z_{V}$ rises from 8 to 9. Further increasing 
$Z_{V}$ to 10 significantly lowers $G$. To explain this fascinating phenomenon, we investigated electronic band structures under finite shear 
strain. As an example, in Fig. \ref{fig2} the band structures (red dotted curves) of NbN and ZrC under a strain of $\epsilon_{12} = 
\epsilon_{21} = 0.1$ are displayed against zero-strain curves. Under shear strain, $p$-$d$ hybridized orbitals split at $\Gamma$-point. 
The energy of the fourth valence band increases dramatically (red thick lines in Fig. \ref{fig2}) in the L-$\Gamma$-K section, and the 
energy of the fifth band decreases. Strain $\epsilon_{4}$ involves shearing metal-N or -C bonds, and the direction-sensitive bonding 
character results in opposite movements of these two bands. For $Z_{V} = 8$ (see band structures of ZrC and HfC in Fig. \ref{fig2}), 
along the L-$\Gamma$-K line, the fourth valence band are nearly fully occupied while the fifth band is empty, so a large $G$ is expected. 
On the other hand, at $Z_{V} = 9$ (NbC, TaC, ZrN, and HfN), valence electrons begin to fill the fifth band, which leads to a negative 
contribution to $G$. However, because the occupation on the fifth band for NbC and TaC is tiny, $G$ of NbC and TaC are very close to those
of ZrC and HfC, respectively. But ZrN and HfN have more electrons filling the fifth valence band around $\Gamma$-point than do NbC and TaC, 
so $G$ of ZrN and HfN are noticeably smaller than those of ZrC and HfC, respectively. A steep decline of $G$ occurs for $Z_{V} = 10$ (NbN and 
TaN) because of substantial filling on the fifth valence band. Note that TaN has the largest $B$ and $G^{\prime}$ among these materials, but
its $G$ is the smallest; and interestingly, TaC have large values of all $B$, $G^{\prime}$ and $G$. 

We fitted three average $G$ points to a simple quadratic form and extrapolated the fitted curve to $Z_{V} = 7$ and 11 (Fig. \ref{fig3}(b)). 
This fitted curve predicts a $G \approx 105$ GPa at $Z_{V} = 7$, and very close to the actual calculated $G = 114$ GPa for YC. YC has smaller 
$G$ than ZrC due to its half filling of the fourth valence band. Our theory predicts a negative $G$ at $Z_{V} = 11$, and indeed we computed 
$C_{44} = -67$ GPa for MoN. Thus MoN is unstable in rocksalt structure because extra electrons fill to the antibonding states, and the 
energy gain from the fourth band is less than the energy loss from the fifth band under shear strain. Previous first-principles calculations 
\cite{chen88} drew the same conclusion. We note that the peak of the fitted curve in Fig. \ref{fig3}(b) is roughly at $Z_{V} = 8.3$, which 
corresponds to full occupation of the fourth valence band without filling on the fifth band, and an optimum $G$ is expected. Non-integer 
$Z_{V}$ can be realized by transition-metal carbonitrides such as ZrC$_x$N$_{1-x}$ or di-transition-metal carbides such as Zr$_x$Nb$_{1-x}$C. 
Our estimation of $Z_{V} \approx 8.3$ for maximum $G$ agrees surprisingly well with experiment \cite{toth71,holleck86,richter96} of $Z_{V} = 8.4$ 
and the {\it ab initio\ } virtual crystal method \cite{jhi99} of $Z_{V}$ in a range of 8.3-8.5. Since both TaC and HfC have large $G$,
we propose that Hf$_x$Ta$_{1-x}$C with $x \approx 0.3$ has the largest shear modulus among these transition-metal carbonitrides or
di-transition-metal carbides in $B_{1}$ structure. Zr$_x$Nb$_{1-x}$C and Ti$_x$V$_{1-x}$C should have comparable $G$ values to 
Hf$_x$Ta$_{1-x}$C, but they have advantages of much less weight.

\begin{table}[tbp]
\caption{Elastic constants (GPa) of selected transition-metal nitrides and carbides with $B_1$ structure. Here theoretical results
were obtained from the DFPT, and experimental data are in parenthesis.
\label{tab2}}
\begin{ruledtabular}
\begin{tabular}{cccc}
system   & $C_{11}$    & $C_{12}$    & $C_{44}$ \\      
\hline	     	     	   
ZrN      & 616 (471\footnotemark[1]) & 117 (88\footnotemark[1])  & 130 (138\footnotemark[1]) \\
ZrC      & 522 (470\footnotemark[2]) & 110 (100\footnotemark[2]) & 160 (160\footnotemark[2]) \\
NbN      & 739 (608\footnotemark[1]) & 161 (134\footnotemark[1]) &  75 (117\footnotemark[1]) \\
NbC      & 667 (620\footnotemark[2]) & 163 (200\footnotemark[2]) & 161 (150\footnotemark[2]) \\
HfN      & 694 (679\footnotemark[1]) & 112 (119\footnotemark[1]) & 135 (150\footnotemark[1]) \\
HfC      & 574 (500\footnotemark[2]) & 107 (114\footnotemark[3]) & 180 (180\footnotemark[2]) \\
TaN      & 783                       & 167                       &  20                       \\
TaC      & 740 (550\footnotemark[2]) & 165 (150\footnotemark[2]) & 176 (190\footnotemark[2]) \\
\end{tabular}
\end{ruledtabular}
\footnotetext[1]{Neutron scattering measurements, Ref.~\onlinecite{chen04}.}
\footnotetext[2]{Estimation from phonon dispersion curves, Ref.~\onlinecite{weber73}.}
\footnotetext[3]{Ultrasonic measurements, Ref.~\onlinecite{brown66}.}
\end{table}

We have carried out first-principles calculations of structural and elastic properties of selected transition-metal nitrides and carbides. 
We computed elastic constants of metals using the density-functional perturbation theory, and our results proved the accuracy of the metric 
tensor approach \cite{hamann04} of the linear response theory of strain perturbation. It has been demonstrated that these crystals have 
similar trends for bulk $B$ and tetragonal shear $G^{\prime}$ moduli, but different trend for shear moduli $G$. $B$ and $G^{\prime}$ are 
associated with strains essentially keeping the crystal symmetry, and all valence bands respond uniformly by moving up or moving down. 
Shear strain $\epsilon_{4}$ lowers the crystal symmetry significantly, and it causes $p$-$d$ orbitals splitting at zone-center and it has 
opposite effects on the fourth and fifth valence bands. Such a systematic study can help clarify the ambiguity rising from different 
experimental methods, and predict new materials with better properties.  

This work was supported by the Office of Naval Research (ONR) Grant number N00014-02-1-0506, the Department of Energy (DOE) Grant 
Nos. DEFG02-02ER4595 and DEFC03-03NA00144, and DOE ASC subcontract B341492 to Caltech DOE W-7405-ENG-48(REC). ZGW thanks the helpful 
discussions with E. J. Walter.

\end{document}